\documentclass[useAMS,usenatbib]{mn2e}

\newcommand{\HI}{{\rm H~{\scriptstyle I}}}

\usepackage{times}
\usepackage{graphicx}
\usepackage{amsmath}
\usepackage{mathabx}
\usepackage{subfigure}
\usepackage{natbib}
\usepackage{txfonts}
\usepackage{journals}

\title[Dust-to-gas ration of GAC]
{Dust-to-gas ratio, $X_{\rm CO}$ factor and CO-dark gas in the Galactic 
anticentre: an observational study}
\author[B.Q. Chen et al.]
{B.-Q. Chen,$^1$\thanks{E-mail:
bchen@pku.edu.cn (BQC); x.liu@pku.edu.cn (XWL).}\thanks{LAMOST Fellow.}
 X.-W. Liu,$^{1,2}$\footnotemark[1]
 H.-B. Yuan,$^2$\footnotemark[2]
 Y. Huang$^1$
 and M.-S. Xiang$^1$
\\
$^{1}$Department of Astronomy, Peking University, Beijing 100871, P.\,R.\,China\\
$^{2}$Kavli Institute for Astronomy and Astrophysics,
Peking University, Beijing 100871, P.\,R.\,China\\
 }

\begin{document}

\date{Accepted ???. Received ???; in original form ???}

\pagerange{\pageref{firstpage}--\pageref{lastpage}} \pubyear{2014}

\maketitle

\label{firstpage}

\begin{abstract}
We investigate the correlation between extinction and H~{\sc i}  and CO emission
at intermediate and high Galactic latitudes ($|b|>10\degr$) within the footprint of the 
Xuyi Schmidt Telescope Photometric Survey of the Galactic anticentre (XSTPS-GAC)
on small and large scales. 
In Paper I (Chen et al. 2014), we present a three-dimensional dust extinction map
within the footprint of XSTPS-GAC, covering a sky area of 
over 6,000\,deg$^2$ at a spatial angular resolution of 6\,arcmin. 
In the current work, the map is combined with data from 
gas tracers, including H~{\sc i}  data from the Galactic Arecibo L-band Feed Array H~{\sc i}  survey
and CO data from the Planck mission, 
to constrain the values of dust-to-gas ratio $DGR=A_V/N({\rm H})$ 
and CO-to-$\rm H_2$ conversion
factor $X_{\rm CO}=N({\rm H_2})/W_{\rm CO}$ 
for the entire GAC footprint excluding the Galactic plane, 
as well as for selected star-forming regions (such
as the Orion, Taurus and Perseus clouds) and 
a region of diffuse gas in the northern Galactic hemisphere. 
For the whole GAC footprint, we find $DGR=(4.15\pm0.01) \times 10^{-22}$\,$\rm mag\,cm^{2}$ 
and  $X_{\rm CO}=(1.72 \pm 0.03) \times 10^{20}$\,$\rm cm^{-2}\,(K\,km\,s^{-1})^{-1}$. 
We have also investigated the distribution of ``CO-dark" gas (DG)
within the footprint of  GAC and found a linear correlation between 
the DG column density and the $V$-band extinction:
$N({\rm DG}) \simeq 2.2 \times 10^{21} (A_V - A^{c}_{V})\,\rm cm^{-2}$.
The mass fraction of DG is 
found to be $f_{\rm DG}\sim 0.55$ toward  the Galactic anticentre, which is respectively about 23 
and 124 per cent of the atomic and CO-traced molecular gas in the same region. This result is consistent with 
the theoretical work of Papadopoulos et al. but much larger than that 
expected in the $\rm H_2$ cloud models by Wolfire et al.

\end{abstract}


\section{Introduction}
The interstellar medium (ISM) makes up between 10 to 15 per cent of 
baryonic  mass of the Milky Way. About 99 per cent of the material 
is in gas form and the rest is in solid dust grains. 
Observation of the interstellar dust and gas is a 
primary tools to trace the structure and distribution of molecule clouds,
where the star formation takes place. The dominant molecular gas
in the ISM is H$_2$,  a homonuclear diatomic molecule with no permanent dipole moment 
and allowed dipole rotational transitions. 
{The most easily excited $\rm H_2$ transition is the quadrupole $S(0)$:~$J_u \rightarrow J_l=2\rightarrow0$, 
which has an excitation energy  $\Delta E_{20}/k \sim 510$\,K, more than an order of magnitude higher than 
the typical kinetic temperature $T_{\rm kin} \sim 10$--50\,K of  the ISM.}
  As a consequence, 
H$_2$ is hard to excite and observe, and 
one has  to trace the molecular gas and determine the
 H$_2$ column density, $N(\rm H_2)$, by other indirect methods.
Two of the most commonly used  tracers  are CO gas
emission and dust extinction.

{Heteronuclear diatomic molecule CO is the most abundant molecule in the ISM after H$_2$. 
The $^{12}$CO $J=1 \rightarrow 0$ transition has a $\Delta E_{10}/k$ about 5.5\,K and can be 
easily excited even in cold molecular clouds and observed from the ground. 
As a consequence, CO has become the primary tracer of the interstellar molecular gas. The 
molecular abundance ratio ${\rm CO}/ {\rm H_2}$ is about $10^{-4}$.}
The  CO ($J=$1$\rightarrow$0) survey of \citet{Dame2001} is the most complete survey
which covers the Milky Way of Galactic latitudes $|b| \leq 30\degr$.  There are
a large number of other CO surveys covering small sky regions around the individual molecular
clouds, such as the Orion and the Monoceros clouds \citep{Hartmann1998, Magnani2000,Wilson2005}.
\citet{Planck2013b} have extracted the full-sky CO 
emission in three lines, the $J=$1$\rightarrow$0, $J=$2$\rightarrow$1
and $J=$3$\rightarrow$2 lines at 115, 230, and 345\,GHz from the Planck 
data using a component separation method.
The $^{12}$CO($J=$1$\rightarrow$0)
emission at wavelength $\lambda=$2.6\,mm is then utilized to determine the 
total column density of molecular gas assuming  a  CO-to-$\rm H_2$
conversion factor, the so-called $X$ factor, which relates 
the surface brightness of CO emission  and the column density of molecular 
gas, $X_{\rm CO}= N({\rm H_2})/W_{\rm CO}$ (cf. \citealt{Bolatto2013} for a review),
where $W_{\rm CO}$ is the velocity-integrated brightness temperature of 
$^{12}$CO $J=1\rightarrow0$.  The $X_{\rm CO}$ factor works best for cloud ensemble averages.
The minimum averaging scale would be that of a single Giant Molecular Cloud (GMC), 
$\sim$10--100\,pc \citep{Dickman1986, Young1991, Bryant1996, Regan2000, Papadopoulos2002}. 
The most quoted value of $X$ factor for the  Milky Way
is $X_{\rm CO} \sim 2 \times 10^{20} \rm cm^{-2}\,(K\,km\,s^{-1})^{-1}$ 
\citep{Frerking1982,Dame2001,Lombardi2006}.

The visual  interstellar extinction $A_V$  has been found to have a linear relationship with the total 
column density of hydrogen nuclei $N(\rm H)$. The relationship is a 
cornerstone of the modern astrophysics. 
The classic value of the constant of proportionality, i.e. the 
dust-to-gas ratio, $DGR=A_V/N({\rm H})$, 
derived from the measurements of the ultraviolet (UV)  absorption of 
the H~{\sc i} Ly$\rm \alpha$ line and the $\rm H_2$ Lyman band  
toward  a small number of early-type stars with  the Copernicus satellite 
\citep{Savage1977, Bohlin1978}, is $DGR=5.34\times10^{-22} \rm \,mag\,cm^{2}$.
More  recently, \citet{Liszt2014a} and \citet{Liszt2014b} 
trace $N(\rm \HI)$ and $E(B - V )$ across the sky of 
Galactic latitudes $|b|$ = 20 -- 60\degr. They  consider data only of relatively low
 extinction,  0.015 $< E(B - V ) <0.075$\,mag, for which  the hydrogen should mostly 
be in the form of neutral atoms and thus the corrections of $N(\rm \HI)$ for 
saturation and $\rm H_2$ formation are unlikely important. From the data, they obtain a
$DGR$ value (3.73$\times10^{-22} \rm \,mag\,cm^{2}$) that is smaller compared to the reference
value. The column density of atomic hydrogen, $N(\rm \HI)$ can be determined by measuring 
the H~{\sc i} 21\,cm emission. Thus if one can accurately determine the extinction and value of $DGR$, then 
the H$_2$ column density
can be derived using the relation,  $N({\rm H_2})=A_V/DGR-N(\rm \HI)$.
There are several successful large H~{\sc i} 21\,cm emission surveys of the Galaxy,
including the Leiden-Argentine-Bonn (LAB) full-sky survey \citep{Kalberla2005},
the international Galactic Plane Survey (IGPS, \citealt{Taylor2003}),
the Galactic All Sky Survey (GASS, \citealt{McClure2005}) in the southern hemisphere,
 and the Galactic Arecibo L-band Feed Array H~{\sc i}  survey (GALFA-HI, 
\citealt{Peek2011}) targeting both the Galactic plane and high latitude regions.

Dust grains can be traced by extinction in the visible and in the near-infrared (IR).
\citet{Chen2014} estimate values of extinction and distance to over
13 million stars catalogued by the Xuyi 1.04/1.20\,m Schmidt Telescope 
photometric survey of the Galactic anticentre
(XSTPS-GAC; \citealt{Zhang2013,Zhang2014,Liu2013}). They derive the extinction by 
 fitting the spectral energy 
distribution (SED) for a sample of 30
million stars that have  supplementary IR photometry from the Two Micron All Sky Survey (2MASS; 
\citealt{Skrutskie1997}) and the
Wide-Field Infrared Survey Explorer (WISE, \citealt{Wright2010}).
Together with the
photometric distances deduced from the dereddened XSTPS-GAC optical
photometry, \citet{Chen2014}
construct  a 3D $r$-band extinction map within  the
GAC footprint at a spatial resolution, depending on the latitude,
that varies  between 3 -- 9\,arcmin. 
The 3D map  of \citet{Chen2014} has also been used to construct a 2D extinction
map, integrated  to a distance of 4\,kpc, approximately the maximum 
distance that  dwarf stars in the sample can trace  
with sufficiently high 
photometric accuracy. The dust extinction within this distance is well constrained.
The map has an angular resolution of 6\,arcmin and is also publicly  
available\footnote{http://162.105.156.249/site/
Photometric-Extinctions-and-Distances.}.
In the current work, this 2D extinction map is used to investigate the 
dust-to-gas ratio and the $X$ factor  within the footprint of XSTPS-GAC.

There is a number of well known large star-forming regions within the footprint of XSTPS-GAC, such as
 the Perseus, Taurus and Orion
clouds. They form parts of the Gould Belt. 
Those clouds are  in the southern Galactic plane, located at 
relatively small distances ($d<800$\,pc; \citealt{Chen2014}). 
We study the correlation between the interstellar dust and gas for these regions. 
Within these regions there exist diffuse, non self-gravitating clouds 
as well as dense, massive, self-gravitating, star-forming clouds like the Orion.
Several studies on the dust and gas correlation have been carried out for those individual clouds. 
\citet{Pinda2008a}, \citet{Goldsmith2008} and \citet{Pinda2010} study the Taurus molecular cloud, while \citet{Pinda2008b}, 
\citet{Lee2012} and  \citet{Lee2014} focus on the Perseus cloud and \citet{Digel1999},
\citet{Ackermann2012} and \citet{Ripple2013} on the Orion cloud.  

One can measure the amount and distribution of the atomic and molecular gas as well as
the dust grains by combining various tracers. In this paper, we combine the dust extinction
together with gas distribution traced by H~{\sc i}  data from the GALFA-HI \citep{Peek2011} 
and CO data from the Planck mission \citep{Planck2013b},
to constrain the values of $DGR$ and $X_{\rm CO}$ at a high angular 
resolution of 6\,arcmin. 
{ The existence of molecular
gas of low CO abundance has been noted previously
in theoretical models \citep{vanDishoeck1988,Papadopoulos2002},
in observations of diffuse gas and high-latitude clouds
\citep{Lada1988} and in observations of irregular galaxies \citep{Madden1997}.
The values of $DGR$ and $X_{\rm CO}$ determined in the current work allow us to further
 trace this so-called ``CO-dark gas" (DG; \citealt{Reach1994, 
Meyerdierks1996, Grenier2005, Abdo2010}) at both small and large scales.}
We are not the first one on this topic of course. For example, \citet{Planck2011b} 
has obtained an all-sky map
of dust optical depth and compared it with the observed distribution of gas column density.
The results are used to  constrain the correlation between the dust and gas,
$(\tau _{\rm D}/N_{\rm H})^{ref}$, as well as the values of  $X_{\rm CO}$, and to construct a 
distribution map of  DG at high and intermediate Galactic latitudes
($|b|>10\degr$). \citet{Paradis2012}  carry out a similar study using the extinction data deduced from
the colour excess of the 2MASS photometry instead of those derived from the
far-IR optical depth. Both the work of  \citet{Paradis2012} and \citet{Planck2011b} are 
limited by the angular resolution of  the LAB 21\,cm H~{\sc i}  
emission survey \citep{Kalberla2005}, which is 36\,arcmin. In the current work we have
adopted data from the GALFA-HI survey. As a consequence,  the angular resolution 
has been improved significantly.
 
The paper is structured as the following. In Section\,2 we present the 
2D extinction map integrated to a distance of 4\,kpc. Section\,3 describes the data of gas
tracers used in the current  analysis.
The correlations between extinction and 
gas content  are investigated in Section\,4 
with the main results  discussed in Section\,5.
We summarize in Section\,6.

\section{The extinction map}
The photometric extinction and distance catalogue of \citet{Chen2014} 
is first used to construct a 2D extinction map 
integrated to a distance of 4\,kpc for the footprint of XSTPS-GAC. The catalogue contains 
more than 13 million stars  with
estimates of  $r$-band extinction and photometric distance.
To ensure good photometric accuracy, only objects flagged with a 
quality flag `A', i.e. with average photometric errors less than 0.05\,mag for  all 
the eight bands used to derive the extinction and distance 
($g,~r$, and $i$ from the XSTPS-GAC, $J,~H$ and $K_{\rm S}$ 
from the 2MASS, and $W1$ and $W2$ from the WISE), are included  (Sample\,A of \citealt{Chen2014}). 
Possible candidates of giants, flagged as `G', are
excluded because their distances may have been grossly underestimated. 
This leaves just over 7 million stars and they are divided into 
sub-fields of size 6 $\times$ 6\,$\rm arcmin^2$. 
For each sub-field, a sliding window of depth 450\,pc, stepping by steps of
150\,pc, is used to obtain the median value of extinction for each distance bin. 
The extinction as a function of distance, $A_r^i(d)=f_i(d)$, where $i$ is the
index of sub-fields,  is then obtained 
after applying a boxcar smoothing of width 450\,pc.
The extinction versus  distance relations  of all sub-fields (sightlines) are
used to construct a 3D extinction map, which is then integrated 
out to a distance of $d$ = 4\,kpc to yield a 2D extinction map to be used in the current analysis.
 Most of the sample stars are within a distance of
4\,kpc and the extinction out to this distance is well 
constrained. The 4\,kpc integrated map can be treated as 
a lower  limit of extinction map along the sightlines. 

\begin{figure}
\centering
   \includegraphics[width=0.48\textwidth]{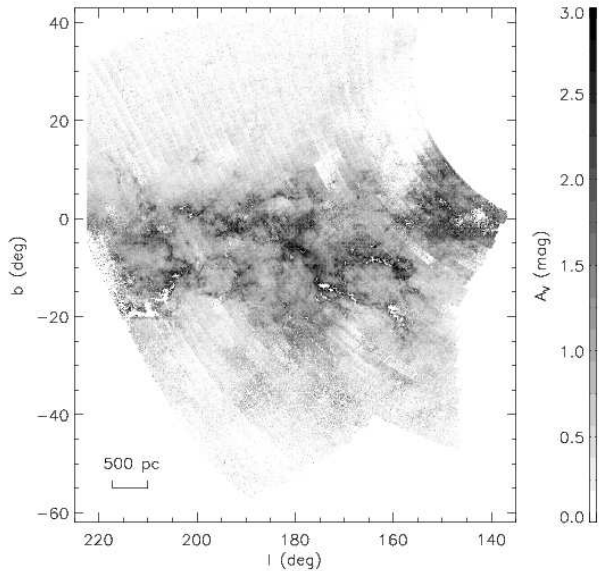}
\caption{A 2D extinction map, integrated to a distance of 4\,kpc for  the 
footprint of XSTPS-GAC. The map has an angular
resolution of 6\,arcmin. } 
\label{extinmap}
\end{figure}

Fig.~\ref{extinmap} presents the 4\,kpc dust 
extinction map thus derived. It has a spatial resolution of  
6\,arcmin. The resolution is chosen to ensure 
that most of the subfields have a sufficient number of stars ($> 10$) to derive a robust 
relation of extinction as a function of distance. For a small fraction of 
sub-fields (less than 1.5 per cent) of high Galactic latitudes or suffering from high extinction, 
there are not enough numbers of  stars of high photometry accuracies
available. For those fields,  we have adopted some stars of lower accuracy
from the catalogue (Sample\,C of
\citealt{Chen2014}). Those sub-fields are flagged by `C'.   
The $V$-band extinction $A_V$ is converted from 
$A_r$ using the extinction law of  \citet{Yuan2013a}, $A_V=1.172A_r$, assuming a 
total-to-selective extinction ratio $R_V=3.1$. 
The uncertainties of the map are estimated from the stars located within the distance
bin of width 450\,pc centered at 4\,kpc and in a width of 450\,pc. The typical uncertainties are 
 0.18\,mag in $A_V$, or about 0.06\,mag in $E(B-V)$. 
The uncertainties are mainly determined by the 
 stellar density and depth of the sample. Under favorable conditions, 
the uncertainties  could be as low as
0.04\,mag in $A_V$. Some ``stripe-like'' artifacts 
can be seen on  the map. They are  
caused by the nightly varying survey depth and photometric accuracy of
the XSTPS-GAC. A  notable empty patch 
around $l \sim$160$\degr$ and $b \sim$20$\degr$ is
due to the abnormally large photometric errors of XSTPS-GAC in that area.

\begin{figure}
\centering
   \includegraphics[width=0.48\textwidth]{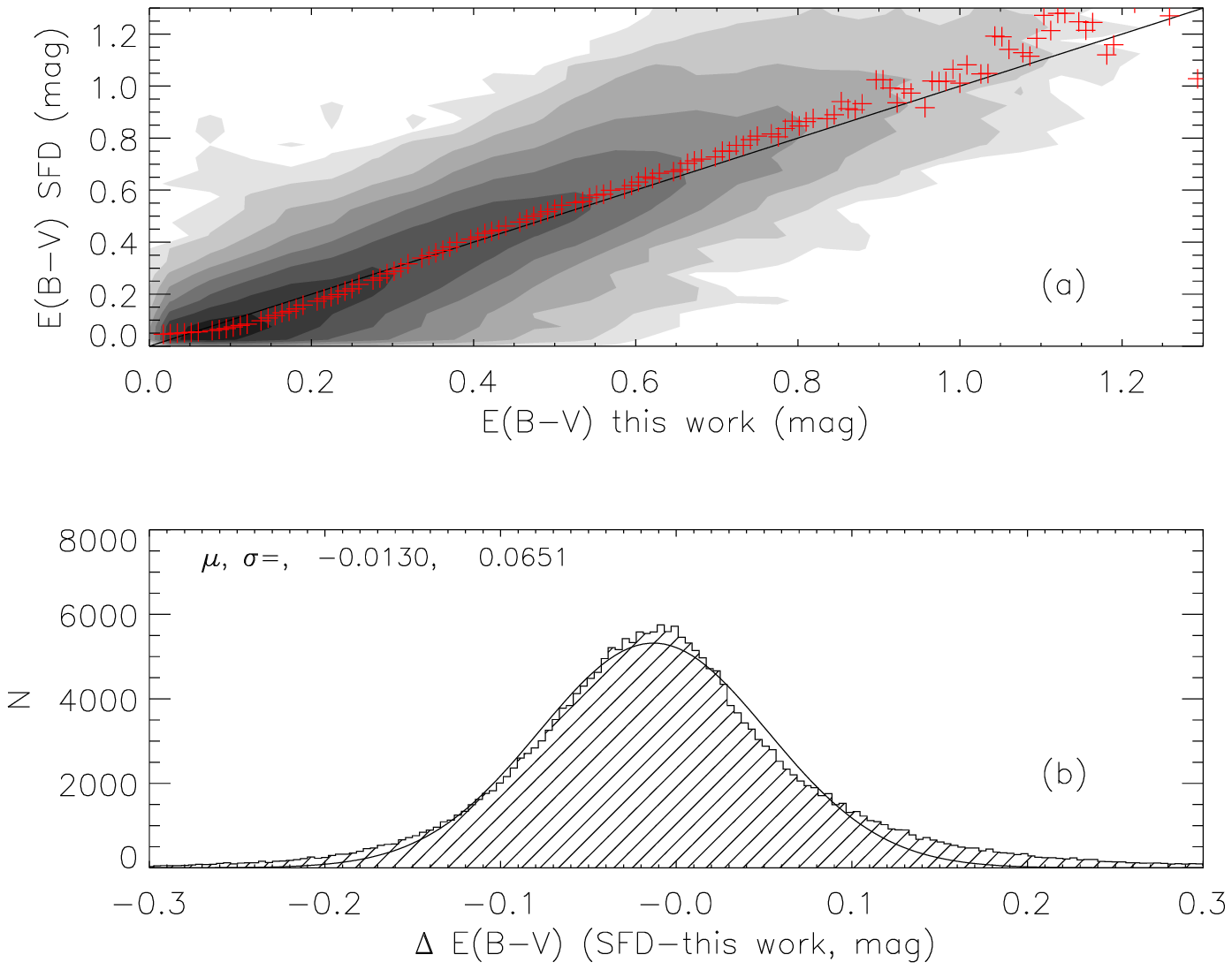}
   \includegraphics[width=0.48\textwidth]{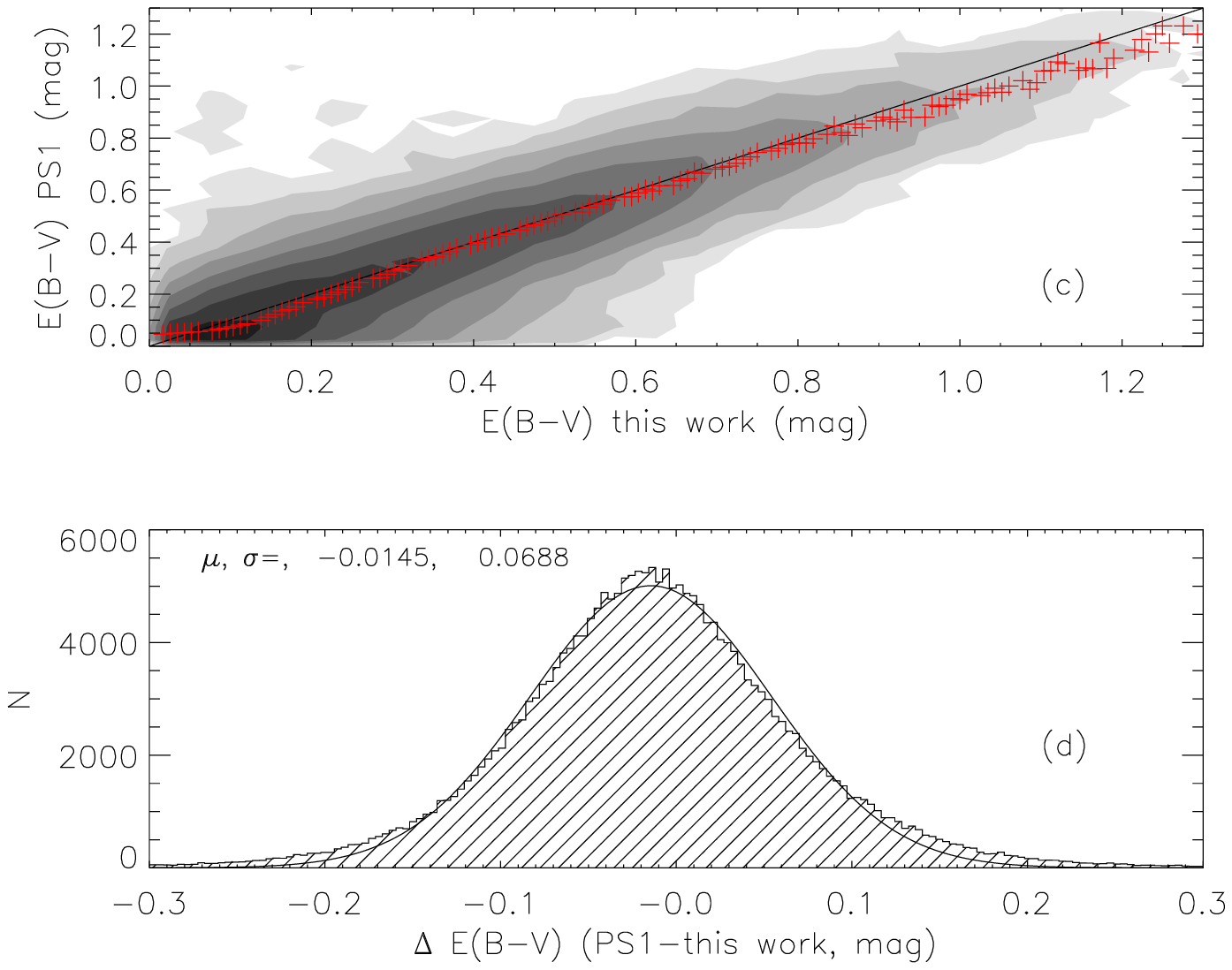}
\caption{Comparisons of extinction values as given by our  map and those 
given by the SFD map (top two panels), and by the map of Sclafly et al. (2014; lower two panels).
The red pluses in the first and third panels represent  median values in the individual  bins. A 
black straight line denoting complete equality is also overplotted to guide the eyes.
 In the second and fourth panels, the black curve is a Gaussian fit to the distribution of differences of values.}
\label{extincomp}
\end{figure}

Chen et al. (2014) have compared their Sample\,A median extinction map 
with those of  \citet{Schlegel1998} and 
\citet{Froebrich2007}, and found some systematic
offsets amongst them. Parts of the differences are clearly caused by  
the difference in distance traced by the three maps [2--3\,kpc for Sample\,A
median extinction map of Chen et al., essentially infinite for  the SFD map and about 1\,kpc 
for that of  \citet{Froebrich2007}, respectively]. The 4\,kpc extinction
map presented here should  have included almost all the dust extinction alone the sightlines
for sub-fields of intermediate and high Galactic latitudes ($|b| > 10\degr$), thus should be directly 
comparable with that of  SFD. In addition, it should also be comparable with the 
4.5\,kpc reddening map newly derived from the PanSTARRS \citep{Kaiser2010}
photometry by  \citet{Schlafly2014}. Fig.~\ref{extincomp} compares values of 
 extinction from our 4\,kpc extinction map with  those of SFD and \citet{Schlafly2014} for 
sightlines of Galactic latitude $|b|>10\degr$. Values of 
$A_V$ are converted to $E(B-V)$ assuming $R_V=3.1$. Except for 
some outliers, our values agrees well with both those of SFD and \citet{Schlafly2014}
for $E(B-V) < 0.8$\,mag.
In deriving the extinction, both our analyses and those of \citet{Schlafly2014}
have  made  the assumption that all 
 reddening values should be no less than zero. The assumption could
introduce some biases, in particular for sightlines of very 
low, close to zero, extinction. In principle, both the best
SED fit algorithm  of \citet{Chen2014} as well as 
the Bayesian approach  of \citet{Green2014} could lead to 
 systematically over-estimated  values of extinction  for regions of  low 
extinction \citep{Berry2012,Chen2014,Schlafly2014,Green2014}.
Photometric errors obviously contribute to the uncertainties. Other potential sources of error
 include  the  colour-reddening degeneracy. Those systematics might be 
the cause of the significant non-Gaussianity seen in the 
distribution of differences of values as given by our map and those of SFD (Fig.~2). 
By contrast, the distribution of differences of values   between ours and those of \citet{Schlafly2014} 
follows closely a Gaussian one, reflecting the fact that both use 
a similar method to estimate the extinction.

For high extinction regions [$E(B-V)>0.8$\,mag], SFD reddening 
values are systematically higher than 
ours while those of \citet{Schlafly2014} map are lower. 
Most of those highly reddened regions for which our estimates of 
reddening are significant smaller than given by the SFD map are actually 
from low Galactic latitudes. In fact, if we restrict the plots of Fig.\,2 to absolute 
latitudes higher than 20$^\circ$ ($|b| > 20^\circ$), then the systematic differences 
between ours and the SFD values disappear. Thus it is very likely, for those highly reddened, 
low latitude regions, the integration distance of 4\,kpc used to derive our current 
2D extinction map is not deep enough. 
The systematic lower values given by the map of \citet{Schlafly2014} 
compared to ours for those highly extincted regions 
 is much more difficult to understand, as both analyses 
use similar methods, although the results are  based on different data set.
The  integration of the  Schlafly et al. map is 500\,pc further
than ours. But instead of giving higher values, the map of Schlafly et al. actually 
yields lower values compared to ours. 
\citet{Schlafly2014} notice that when their map is compared to that of SFD, 
their map yields systematic lower values even for regions of latitudes  
$|b|>30\degr$. It is possible that 
\citet{Schlafly2014} may have systematically underestimated the 
reddening values for some of the highly
extincted regions. For the whole XSTPS-GAC footprint of $|b|>10\degr$, 
our map gives on average only marginally larger values compared to those of SFD and 
\citet{Schlafly2014}, by 0.013 and 0.015\,mag,
respectively. The scatters of differences between the values are about 0.06\,mag 
in $E(B-V)$, comparable to the typical
uncertainties estimated for  our map (0.18\,mag in $A_V$). 

\section{Gas tracers}

The interstellar gas can either be atomic, molecular or ionized. 
Those three phases are commonly traced by the H~{\sc i}  21\,cm emission, 
the molecular CO pure rotational 
transitions and by the H~{\sc i} $\rm H \alpha$ recombination line, respectively. 
The ionized gas can also be traced by the free-free emission of thermal electrons. 
The contribution of the ionized gas,
which are mostly restricted to the individual, isolated  H~{\sc ii} regions, 
 to the total gas content is neglected in the current study. 
Only most recently published data that  have the highest angular resolutions 
and best signal-to-noise ratios (SNRs) are employed in this work.

\subsection{H~{\sc i}  data}

We use the H~{\sc i}  data from the GALFA-HI survey  to trace the 
content of atomic gas. The
GALFA-HI  survey uses Arecibo L-band Feed Array, 
a seven-beam array of receivers mounted at the focal 
plane of the 305\,m Arecibo telescope, to map the H~{\sc i}  
21\,cm emission in the Galaxy 
\citep{Peek2011}. The survey aims to cover both the Galactic 
plane as well as the  high Galactic  latitudes with an angular resolution of 3.4\,arcmin.
The GALFA-HI first data release (DR1) has been presented by \citet{Peek2011}. 
The DR1 data covers 7520\,deg$^2$ of the sky, produced from 3046\,hr worth
of data obtained for 12\,individual projects. 
The local standard of rest (LSR) velocity of the emission 
ranges from $-700$ to $+700$\,$\rm km\,s^{-1}$.
The rms noise in a 1\,km\,$\rm s^{-1}$ 
channel ranges from 140\,mK to 60\,mK, with a median of 80\,mK. 
The DR1 data can be downloaded from https://purcell.ssl.berkeley.edu/. 

\begin{figure*}
\centering
   \includegraphics[width=0.48\textwidth]{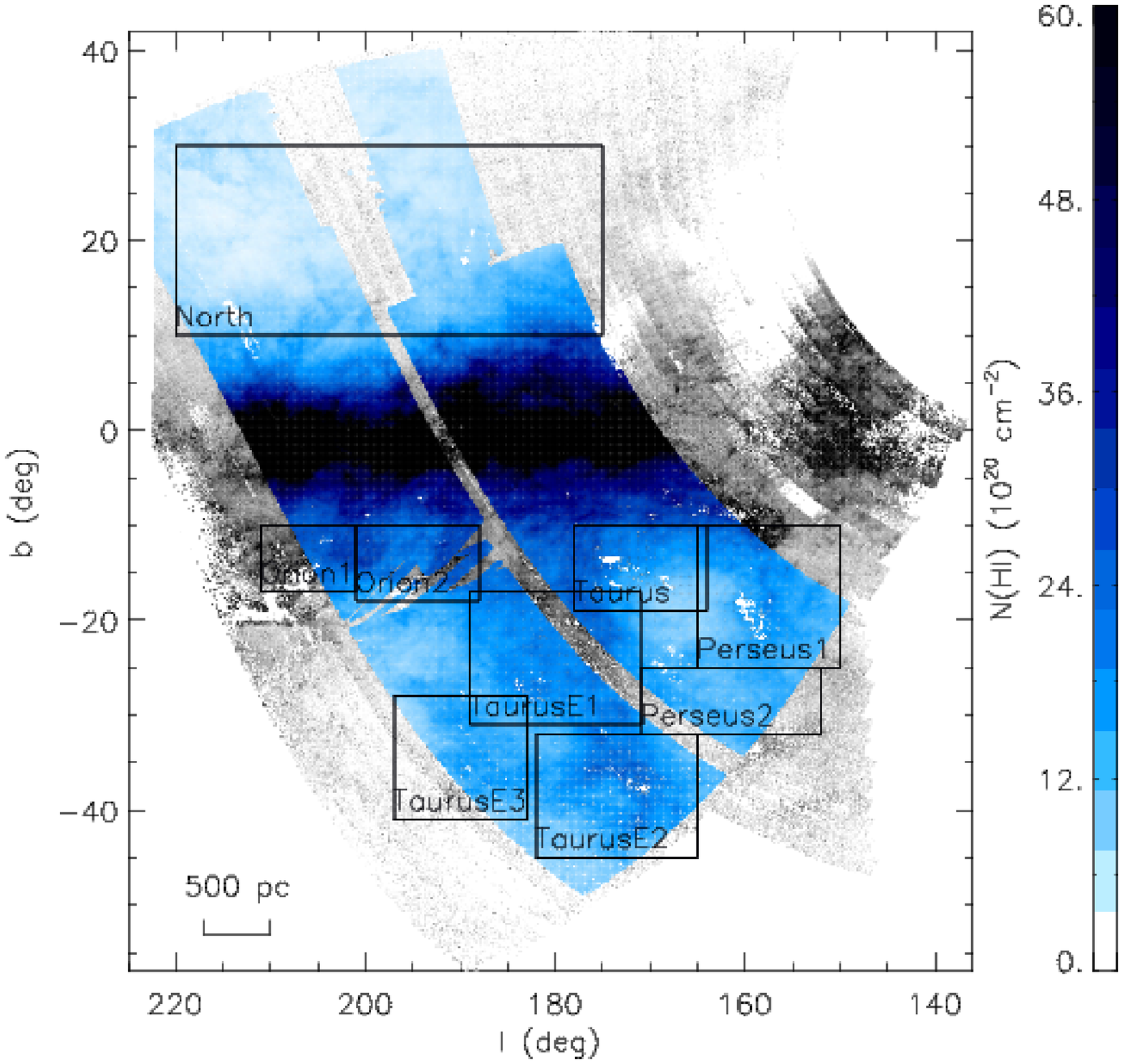}
   \includegraphics[width=0.48\textwidth]{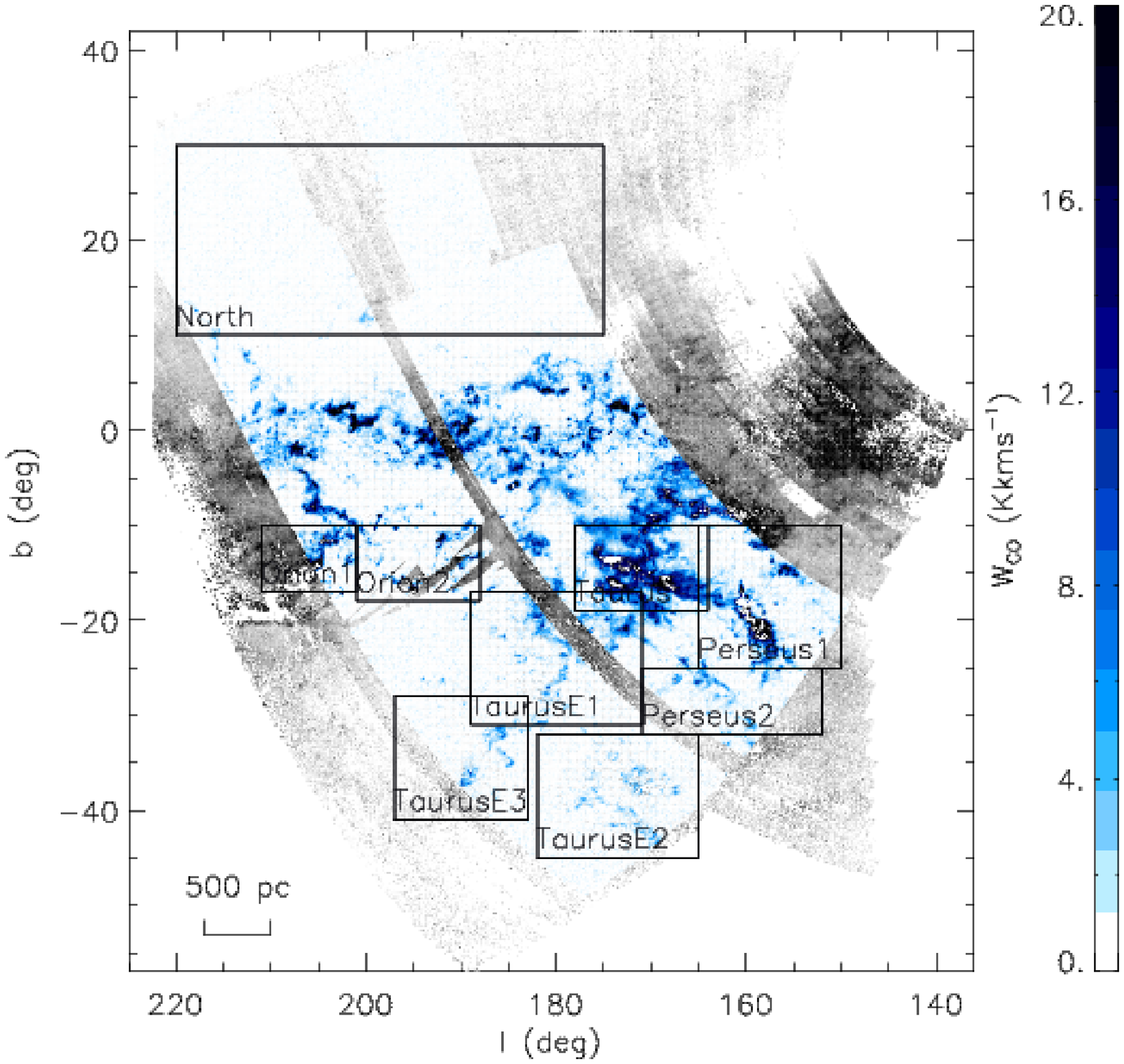}
\caption{Contours of $N{\rm (\HI)}$ (left panel) and 
$W_{\rm CO}$  (right panel) in the footprint of XSTPS-GAC. The grey-scale background image illustrates
the 4\,kpc extinction map. The squares denote 
different regions analyzed in the current work. {The maps have an angular
resolution of 6\,arcmin.}  }
\label{hicomap}
\end{figure*}

The GALFA-HI DR1 data does not cover the entire  
footprint of XSTPS-GAC, but only the region of 
$\rm 3h~<~RA~<~9h$ and $0~<~{\rm Dec} ~<~ 39$\,deg. 
This unfortunately restricts  the sky coverage of the current work.
We use the data of velocity resolution  0.8\,$\rm km\,s^{-1}$. 
The H~{\sc i} column density $N({\rm \HI})$ is derived from the 
profile of brightness temperature after applying a small 
correction for the optical depth effects assuming  a spin
temperature of 145\,K, as adopted by \citet{Liszt2014a,Liszt2014b}. 
The temperature  corresponds to the mean ratio of H~{\sc i} 21\,cm emission and 
absorption \citep{Liszt2010} measured by  the Millennium H~{\sc i}  
Emission-Absorption Survey of \citet{Heiles2003}. 
Values of  $N(\rm \HI)$ derived for the XSTPS-GAC footprint range from 
1.5 $\times 10^{20}$ to 1.2 $\times 10^{22}\, \rm cm^{-2}$, 
with a median of 1.4 $\times 10^{21}\, \rm cm^{-2}$. 
The distribution of  $N(\rm \HI)$ thus deduced is shown in the left panel of 
Fig.~\ref{hicomap}. 

\subsection{CO data}

We use the CO-integrated intensity image produced by \citet{Planck2013b},
which is based on the Planck first 15.5\,month survey data 
(two full-sky scans) and uses the full-sky maps of the nine 
Planck frequency bands, and also the 100, 217 and 353\,GHz 
full-sky bolometer maps. \citet{Planck2013b} provide all sky maps of the 
velocity-integrated  emission of the CO $J=1 \rightarrow 0,~J=2 \rightarrow 1$ and 
$J=3 \rightarrow 2$ transitions. They produce
three types of  maps. We choose the { Type~}3 map 
generated from the multi-line approach. The approach assumes fixed CO line
ratios in order to achieve the highest possible signal-to-noise 
ratios, and has  an angular resolution of $\sim 5.5\, \rm arcmin$. 
The standard deviation of {the Type~}3 CO(1$\rightarrow$0) map, 
when used at an angular  resolution of 15\,arcmin, {is 
typically 0.16\,$\rm K$, compared to 
1.77 and  0.45\,$\rm K$ for the {Type~}1 and 2 maps, respectively.
For comparison,  the CO survey of \citet{Dame2011} has a  typical uncertainty of 
0.6\,$\rm K$.}
We choose the Planck { Type~}3 CO (1$\rightarrow$0) map because of its high 
angular resolution and high SNR. {Type~}3 map is also 
more suitable for intermediate and high-Galactic 
latitudes (cf. \citealt{Planck2013b}  for more detail).
There is some $^{13}$CO contamination in the Planck CO(1$\rightarrow$0) map. 
To correct for this, we have divided the data by a factor of 1.16 \citep{Planck2013b}. 

\citet{Dame2001} combine  data of  CO surveys of the Galactic plane and local molecular 
clouds and produce a composite map of an angular 
resolution of 8.4\,arcmin. The data were obtained with the 1.2\,m telescope of 
the Harvard-Smithsonian Center for Astrophysics. The  data
published by \citet{Dame2001} are mainly for the Galactic plane. 
Data of high Galactic latitudes ($|b|>20\degr$) are not available. In this paper, the data of 
\citet{Dame2001} are only used as a reference to check 
the data we actually use. 

The Planck {Type~}3 CO-integrated intensity image is presented in the right panel of 
Fig.~\ref{hicomap}. We only show the footprint where data of  $A_V$, $N$(H~{\sc i} ) and 
$W_{\rm CO}$ are all  available. It shows excellent  agreement with the dust  extinction map.
In this paper, we first bring all input data sets to a common angular 
resolution determined by the lowest angular resolution data set, which is 
6\,arcmin of the extinction map. This is done by performing a Gaussian kernel smooth. 
Thanks to the high resolution, we can analysis the correlation between 
dust and gas at  intermediate Galactic latitudes of the GAC
for the individual molecular clouds,
such as the Orion, Taurus and Perseus cloud. We divide the clouds into 
eight different regions. Their positions and ranges on the sky are marked by  the
squares in Fig.~\ref{hicomap}, together with a region of diffuse gas without 
any obvious molecular clouds in the northern Galactic hemisphere, noted as ``North'' in the current work.

\section{Extinction and gas correlation}


The observed visual extinction $A_V$ can be modelled by 
contributions from the atomic gas  and the
CO-traced molecular gas, as
\begin{equation}
A^{mod}_{V}=DGR(N({\HI})+2X_{\rm CO}W_{\rm CO})+A^0_V,
\end{equation} 
where  $A^0_V$ is a constant  to 
account for possible measurement  offset of the extinction map 
and/or possible contribution to the extinction from the ionized gas. 
The total gas column density is given by, $N({\rm H})=N({\HI})+2N({\rm H_2})$.
We constrain the three free parameters, $DGR,~X_{\rm CO}$ and $A^0_V$, 
using the visual extinction, H~{\sc i}  column density and 
CO integrated line intensity maps described in the last Section.
Considering  the high resolution of the maps and the large 
uncertainties of extinction  for low extinction regions, 
we have adopted a fitting strategy similar  to that 
of \citet{Paradis2012} . The fitting process can be  described as following.

\begin{enumerate}
\item We first assume that there is no DG in all subfields (pixels). We then select pixels where
the CO emission is low ($W_{\rm CO}< 0.2$\,K\,km\,$\rm s^{-1}$). 
The  column density of molecules of those pixels
$N(\rm H_2)$ must be less than  $2.0 \times 10^{19}\, \rm cm^{-2}$ if one assumes
that $X_{\rm CO}=1 \times 10^{20}\, \rm cm^{-2}\,(K\,km\,s^{-1})^{-1}$. Given 
that the minimum column density of atomic gas within the 
XSTPS-GAC footprint is $1.5 \times 10^{20}\, \rm cm^{-2}$ (see Section~3.1), the
gas column density of those pixels can be considered as 
being contributed by atomic gas only [$N$(H)=$N$(H~{\sc i} )]. Under such assumptions, we then derive the 
best-fit values of $DGR$ and $A^0_V$ of those pixels.

\item The best-fit values of  $DGR$ and $A^0_V$ obtained from the last step 
are then applied to all pixels, and this time the value of $X_{\rm CO}$ is fitted as a free parameter.  
{We have smoothed the data to an angular resolution of 2\degr, corresponds to about 140\,pc at a distance of 4\,kpc
and 15\,pc at a distance of 400\,pc (where most of the molecular clouds analyzed in the current work, the Orion,
Taurus and Perseus clouds locate within) in this step to obtain the statistical 
cloud-ensemble-average values of $X_{\rm CO}$ \citep{Dickman1986, Young1991, Bryant1996, 
Papadopoulos2002}.}

\item The above derived values of $DGR$, $A^0_V$ and $X_{\rm CO}$ are applied to Eq.~(1) 
to calculate values of $A^{mod}_V$ of all pixels. A $\chi ^2$ value is then 
calculated as $\chi ^2=\sum_{i} {(A_{V,i} - A^{mod}_{V,i}})$, where  $i$ is the pixel index.

\item The column density of DG is then calculated from  formula,
\begin{equation}
N({\rm DG})=(A_V-A^{mod}_V)/DGR.
\end{equation} 

\item We then go back to the first step, but this time use 
the DG distribution derived from the previous
step. We select pixels that both have low CO emission ($W_{\rm CO} <
0.2$\,K\,km\,$\rm s^{-1}$) and no DG content [$N(\rm DG) 
\leq 0$\,$\rm cm^{-2}$]. The fields are used to update the best-fit 
values of $DGR$ and $A^0_V$ and the process is 
iterated until a minimum $\chi ^2$ value is reached .

\end{enumerate} 


In the current work, we have omitted the Galactic plane which shows very
strong atomic gas emission and where the contribution of ionized gas can not be ignored. 
Thus only data from Galactic latitudes $|b|>10\degr$ are used throughout the paper.
{Since we have excluded the Galactic-plane clouds, those included are 
representative of the diffuse cloud component of the Galaxy.}
Usually a minimum $\chi ^2$ is reached  after the first iteration that assumes there is no DG 
component at all pixels. We list the results
in Table\,1 for both the whole XSTPS-GAC footprint of $|b| > 10^\circ$ and for the individual selected regions. 
Errors of the resultant parameters are derived using a 
bootstrap method. For each  parameter, we apply the algorithm to 1,000 randomly 
selected samples and derive  the parameters of those samples. 
The resultant parameters follow a 
Gaussian distribution. The variance of the Gaussian distribution is taken 
as the error of the corresponding parameter. 

\begin{table*}
 \centering
  \caption{Dust to gas ratio $DGR$ and CO-to-H$_2$ conversion factor $X_{\rm CO}$ derived for individual regions toward the Galactic anticentre.}
  \begin{tabular}{lcrccccccc}
  \hline
  \hline
Region & $DGR$ & $A^0_V$ & $X_{\rm CO}$  & $\dfrac {M_{\rm H}({\rm DG})}{M_{\rm H}({\rm H~{\scriptstyle I}})}$ &
$\dfrac{M_{\rm H}({\rm DG})}{M_{\rm H}({\rm H^{\rm CO}_2})}$ & $f_{\rm DG}$ & $\dfrac {N(\rm DG)}{(A_V-A^c_V)}$ & $A^c_V$ \\
& ($\rm 10^{-22}\,mag\,cm^{2}$) & (mag) & ($\rm 10^{20}\,cm^{-2}\,(K\,km\,s^{-1})^{-1}$)  
&  &  &   & ${\rm 10^{21}\,cm^{-2}\,mag^{-1}}$  & (mag) \\
 \hline
All of $|b|>10\degr$ & $4.15\pm0.01$ & $0.06\pm0.002$ & {1.72$\pm$0.01} 
& {0.23} & {1.24} & {0.55} &             & \\
Orion\,1            & $7.74\pm0.21$ & $-0.75\pm0.05$ & {1.54$\pm$0.01}
& {0.09} & {0.45} & {0.31} & {2.15$\pm$0.07} &{1.10$\pm$0.05} \\
Orion\,2            & $5.53\pm0.08$ & $-0.29\pm0.02$ & {1.34$\pm$0.03}
& {0.10} & {1.43} & {0.59} & {1.69$\pm$0.07} &{0.97$\pm$0.06} \\
Taurus              & $3.65\pm0.10$ & $0.34 \pm0.02$ & {1.38$\pm$0.01}
& {0.22} & {0.31} & {0.24} & {2.75$\pm$0.12} &{0.91$\pm$0.04} \\
Taurus E1         & $6.42\pm0.10$ & $-0.15\pm0.02$ & {0.84$\pm$0.01}
& {0.11} & {1.43} & {0.59} & {2.28$\pm$0.02} &{0.76$\pm$0.04 }\\
Taurus E2         & $5.62\pm0.08$ & $-0.14\pm0.01$ & {1.69$\pm$0.04}
& {0.11} & {4.01} & {0.80} & {1.89$\pm$0.07} &{0.64$\pm$0.05 }\\
Taurus E3         & $6.33\pm0.15$ & $-0.13\pm0.02$ & {1.41$\pm$0.02}
& {0.13} & {3.42} & {0.77} & {2.28$\pm$0.03} &{0.57$\pm$0.04} \\
Perseus\,1         & $5.01\pm0.08$ & $0.10 \pm0.01$ &{1.04$\pm$0.01}
& {0.18} & {0.89} &{ 0.47} & {2.10$\pm$0.04} &{0.59$\pm$0.05} \\
Perseus\,2         & $3.24\pm0.28$ & $0.22 \pm0.03$ &{2.39$\pm$0.01}
& {0.22} & {1.05} & {0.52} & {2.36$\pm$0.04} &{0.55$\pm$0.04} \\
North                 & $1.38\pm0.02$ & $0.20\pm0.002$&
& {0.95} &                      &                    & {2.40$\pm$0.03} &{0.22$\pm$0.04} \\
 \hline
\end{tabular}\\
  \label{ta1}
\end{table*}

\begin{figure}
\centering
   \includegraphics[width=0.48\textwidth]{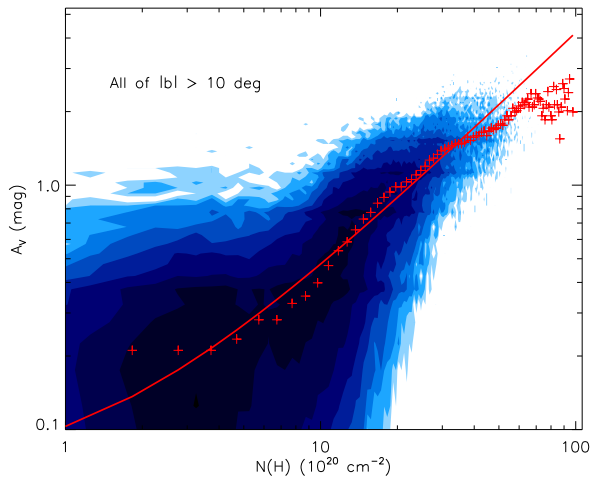}
\caption{Correlation between the visual extinction $A_V$
and the total column density of hydrogen, $N({\rm H})=N(\HI)+2N(\rm H_2)$,
for the whole XSTPS-GAC footprint of $|b|>10\degr$ that excludes the Galactic plane. The 
blue contours  represent the density of pixels on a logarithmic  scale. The 
red pluses are median values for the individual  $N$(H) bins. The red
solid line is  a fit to all the data points.}
\label{dgrall}
\end{figure}

We present in Fig.~\ref{dgrall} the correlation between visual  extinction $A_V$
and total column density of hydrogen, $N({\rm H})=N({\HI})+2X_{\rm CO}W_{\rm CO}$,
for all XSTPS-GAC pixels of $|b|>10\degr$, excluding the Galactic plane. Despite the 
large scatters, in particular for pixels of low extinction, the data exhibit 
a significant linear correlation between the two quantities. The data yield a best-fit value of $DGR$ of $(4.15 \pm 0.01) \times
10^{-22}\, \rm mag\,cm^{2}$. 
{The best-fit value of $X_{\rm CO}$
is found to be $(1.72 \pm 0.01) \times 10^{20}\,\rm 
cm^{-2}\,(K\,km\,s^{-1})^{-1}$.}

\begin{figure*}
\centering
   \includegraphics[width=0.85\textwidth]{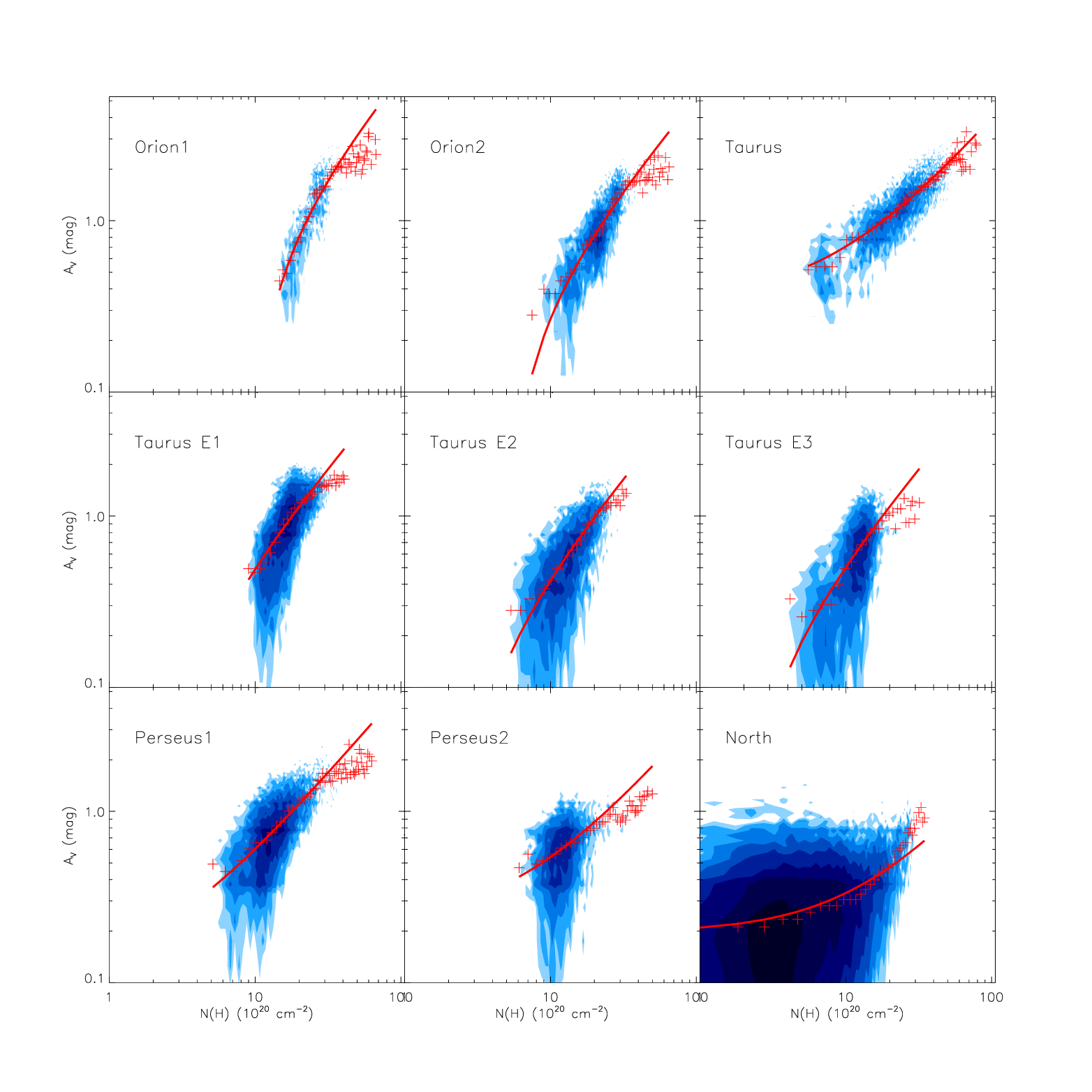}
\caption{Same as Fig.~\ref{dgrall} but for the individual regions. The
name of the region is marked in  each panel.}
\label{dgr}
\end{figure*}

The southern Galactic hemisphere covered by  the XSTPS-GAC survey ($b < -10\degr$) include  several
prominent giant molecular clouds, including the Orion, Perseus, Taurus clouds and the Taurus cloud extension regions. 
We have earmarked eight regions and examine  
the gas and dust correlation. We have also applied our algorithm to a region of diffuse gas in the northern Galactic hemisphere,
earmarked  as ``North''.
The ranges, mean extinction, mean $N{\rm (\HI)}$  and mean
$W_{\rm CO}$ of these regions 
are listed in Table~2.
The positions of those regions are marked by squares in Fig.\ref{hicomap}. 
The CO and extinction maps show remarkably similar 
features, although those features are not so obvious in the H~{\sc i}  map.
For those individual regions, the resultant parameters are also listed in 
Table\,1 and the relevant data showing the correlation between
$A_V$ and $N(\rm H)$ are presented in Fig.~\ref{dgr}.
Clearly, different clouds seem to have different dust and gas properties. The values of $DGR$ 
deduced for the individual regions of molecular clouds range from $3.24 \times 10^{-22}$ to 
$7.74 \times 10^{-22}$\,$\rm mag\,cm^2$. For the diffuse gas region ``North'', the value is
$1.38 \times 10^{-22}$\,$\rm mag\,cm^2$. {The values of $X_{\rm CO}$ for different clouds range from 
$0.84 \times 10^{20}$ to 
$2.39 \times 10^{20}\,\rm cm^{-2}\,(K\,km\,s^{-1})^{-1}$}.
The correlation between $A_V$ and $N(\rm H)$ is quite good for the
individual clouds. In the ``North'' region, 
the lack of high CO content pixels makes it very difficult to fit $X_{\rm CO}$.
The dust extinction in this region  is also low (most of the pixels  have $A_V<1$\,mag), 
leading to a value of $DGR$,  much smaller than those found for the regions  of molecular clouds.

\begin{table*}
 \centering
  \caption{Individual regions studied in the current work.}
  \begin{tabular}{lccccc}
  \hline
  \hline
Region & longitude & latitude & $\overline{A_V}$ (mag) & 
$\overline{N(\HI)}$ ($\rm 10^{20}\,cm^{-2}$) & $\overline{W_{\rm CO}}$ ($\rm K\,km\,s^{-1}$)  \\
 \hline
Orion\,1 & 211\degr$ <l< $ 201\degr &  $-17\degr  <b<   -10$\degr &  1.44 & 23.93 & 1.61 \\
Orion\,2  & 201\degr $ <l< $  188\degr &  $-18\degr    <b<  -10\degr$ & 1.04 & 22.31 & 0.56 \\
Taurus   & 178\degr $ <l< $  164\degr&  $-19\degr  <b<  -10\degr$ & 1.38  & 16.47 & 4.22 \\
Taurus E1  & 189\degr $ <l< $  171\degr & $ -31\degr  <b<  -17\degr$ & 1.02  & 16.78 & 0.75 \\
Taurus E2  & 182\degr $ <l< $ 165\degr&  $-45\degr  <b<  -32\degr$ & 0.70 & 14.62 & 0.12 \\
Taurus E3  & 197\degr $ <l< $ 183\degr & $ -41\degr  <b<  -28\degr$ & 0.66 & 11.96 & 0.16 \\
Perseus\,1  & 165\degr $ <l< $ 150\degr & $ -25\degr  <b<  -10\degr$ & 0.93 & 13.35 & 1.30 \\
Perseus\,2  & 171\degr $ <l< $ 152\degr & $ -32\degr  <b<   -25\degr$ & 0.65 & 11.49 & 0.49 \\
North  & 220\degr $ <l< $ 175\degr& 10\degr$ <b< $ 30\degr& 0.31 & 7.70 & 0.00 \\
 \hline
\end{tabular}\\
  \label{ta1}
\end{table*}

\section{Discussion}

\subsection{Comparison with previous work}

The average value of $DGR$ deduced here for the XSTPS-GAC footprint  
that excludes the Galactic plane, $(4.15\pm 0.01)\times 10^{-22}$\,mag\,cm$^2$,
is in good agreement with the results of previous
work. It is slightly  smaller than the
 most cited value, 5.3$\times 10^{-22}\,\rm mag\,cm^{2}$, derived by 
\citet{Bohlin1978} from the UV and 
optical absorption line measurements toward hot stars. It is
also slightly smaller than values of $(6.0-6.5) \times 10^{-22}\,\rm mag\,cm^{2}$ derived
from the reddening and the Ly$\rm \alpha$ absorption toward early-type 
stars \citep{Shull1985, Diplas1994}. On the other hand, \citet{Liszt2014a} finds an even smaller value
of $3.7 \times 10^{-22}\,\rm mag\,cm^{2}$ from the SFD 
map combined with 21\,cm H~{\sc i}  measurements. The values 
of $DGR$ for the individual cloud regions vary by a factor of two,  
from $3.2$ to $7.7 \times 10^{-22}\,\rm mag\,cm^{2}$.
They are also consistent with values from the previous work. \citet{Paradis2012} obtain an average
value of $5.46 \times 10^{-22}\,\rm mag\,cm^{2}$ for the whole region of the outer Galaxy ($|l|>70\degr$) and values
of $(3.4-8.7) \times 10^{-22}\,\rm mag\,cm^{2}$ for individual regions, 
in good agreement with our results. A variety of factors may 
lead to variations of dust-to-gas ratios amongst  different regions, 
including differences in dust content and metal abundance. 
The dust extinction law of different regions may also differ. \citet{Froebrich2007}
discuss variations of the extinction law amongst individual clouds in the GAC area. In fact,  we notice  
that the value of  $DGR$ decreases with increasing power-law index $\beta$ 
of extinction law in the near-IR. The index $\beta$ is known to correlate  with the dust properties, 
including the grain size distribution \citep{Draine2003}.

The value of CO to H$_2$ conversion factor, $X_{\rm CO}$, determined hitherto differs from one 
analysis to another. {Our best-fit value of $1.72 \times 10^{20}
\, \rm cm^{-2}\,(K\,km\,s^{-1})^{-1}$ for the whole XSTPS-GAC footprint of $|b| > 10^\circ$ is close 
to the Galactic average value \citep{Strong1996}.} The value is
slightly lower but nevertheless quite close to the 
value of $(1.8 \pm 0.03) \times 10^{20}\,\rm 
cm^{-2}\,(K\,km\,s^{-1})^{-1}$, derived  for Galactic regions of $|b|>5\degr$ 
by \citet{Dame2001} based on extinction data from  SFD, H~{\sc i} data 
from the Dwingeloo-Leiden H~{\sc i}  survey  and CO data from \citet{Dame2001}.
 {\citet{Glover2011} demonstrate by numerical simulations that $X_{\rm CO}$
falls off with extinction as $X_{\rm CO}~ \propto ~A^{-3.5}_V$ for clouds of mean extinction 
smaller than 3\,mag. However, our current results do not show that kind of relation. In addition,  our derived values of 
$X_{\rm CO}$ are significantly smaller than predicted by the numerical simulations of \citet{Glover2011}.}
{\citet{Polk1988} consider a simple two-component cloud model which consists of GMC cores and lower
column density (``diffuse'') molecular gas. The conversion ratios for the dense cloud cores and the diffuse
clouds are different, estimated at $\sim4 \times 10^{20}$ and $0.5 \times 10^{20}
\, \rm cm^{-2}\,(K\,km\,s^{-1})^{-1}$, respectively. Thus the measured values of $X_{\rm CO}$
in the current work (around $\sim 1.5 \times 10^{20}
\, \rm cm^{-2}\,(K\,km\,s^{-1})^{-1}$ ) are closer to those of the diffuse clouds of the work
by  \citet{Polk1988} than the dense ones, indicating that the clouds in our sample are dominated
by the diffuse phases. }
\citet{Pinda2008a} find that $X_{\rm CO}$ varies from 0.9 to $3.0
~\times 10^{20} \,\rm cm^{-2}\,(K\,km\,s^{-1})^{-1}$ amongst several  regions 
of the Perseus cloud. \citet{Ackermann2012} find that for the  Orion cloud,
$X_{\rm CO} \sim (1.3 \sim 2.3) \times 10^{20} \,\rm cm^{-2}\,(K\,km\,s^{-1})^{-1}$.
Finally, \citet{Abdo2010} report a significant 
variations ranging from $\sim$0.9$\times$ to  $\sim 1.9 \times 10^{20}\,\rm cm^{-2}\,(K\,km\,s^{-1})^{-1}$, 
from  the Gould Belt to the Perseus arm. 
All those findings  are all in good agreement with our results.

\subsection{Validation of the gas tracers}

\begin{figure}
\centering
   \includegraphics[width=0.45\textwidth]{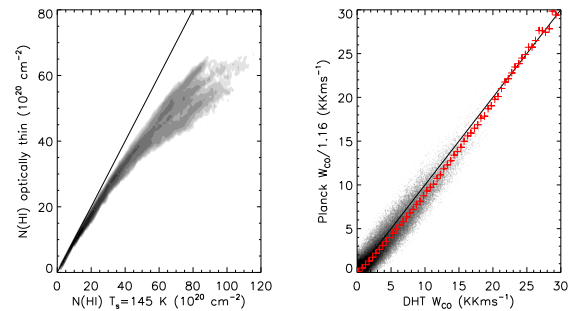}
\caption{{\em Left panel}: Comparison of atomic hydrogen column densities derived assuming a 
constant spin temperature of  $T_s=145$\,K and those deduced assuming  
optically thin emission. {\em Right panel}:
Comparison of the corrected Planck CO data and those of \citet{Dame2001}. 
The red pluses are  median values for the individual  bins and the black
solid lines represent the line of equal.}  
\label{cohi}
\end{figure}

Both \citet{Planck2011b} and \citet{Paradis2012} assume  that
the 21\,cm line is optical thin when calculating the 
column density of atomic gas.
\citet{Fukui2014} infers that the DG in the Galaxy is dominated
by optically thick and cool H~{\sc i} gas, which implies that the average density of
H~{\sc i} is two times higher than that derived assuming  optically-thin in the
local interstellar space.
 In this work we have applied a first-order opacity
correction, assuming a constant $T_s$ for the emission. The need to correct
for the opacity effects in the 21\,cm line profiles ultimately
limits the accuracy of $N({\rm H})$ determinations. We find that for most directions 
toward the GAC but out of the Galactic plane, the
column densities of neutral hydrogen derived are systematically higher  
than those deduced assuming optically thin.
A comparison of H~{\sc i} column densities derived assuming  optically thin and
those assuming  a constant $T_s=145$\,K is presented in the left panel of Fig.~\ref{cohi}. 
The latter values are 1.1--1.3 times higher than  their optically thin counterpart.

In the right panel of Fig.~\ref{cohi}, we compare 
the Planck CO data with those of \citet{Dame2001}. The Planck Type 3 CO  data are  in good 
agreement with those of  \citet{Dame2001}, after correcting for a constant slope of 
1.16. We have investigated the parameters for the overlapping subfields 
using the CO data of \citet{Dame2001} instead of those Planck. 
It yields consistent results very similar to those presented here.

\subsection{The CO dark gas}

There is a clear excess of $A_V$ as given by  the extinction map compared to the simulated 
$A^{mod}_V$ for regions of $N({\rm H}) > 1.2 \times 
10^{21}\,\rm cm^{-2}$ ($A_V>0.6$\,mag, see Fig.~\ref{dgrall}). 
The excess corresponds to the presence of CO dark gas, i.e. DG \citep{Planck2011b,
Paradis2012}. The DG could be interstellar gas that exists in the form of $\rm H_2$
along with C~{\sc i} and C~{\sc ii}, but contains little or no CO.
 The column density of DG, $N(\rm DG)$, can be determined using Eq.\,(2).
The total mass of hydrogen of the DG component, $M(\rm DG)$, the mass of 
atomic hydrogen, $M(\HI)$, and the mass of the  CO-traced molecular gas, 
$M(\rm H^{\rm CO}_2)$, can be  estimated
following  \citet{Planck2011b} and \citet{Paradis2012}: 
$M({\rm DG/ \HI /H^{\rm CO}_2}) \propto \sum N({\rm DG/ \HI /H^{\rm CO}_2})$.
We have  assumed that all gas component have  the same spatial extent. 
{We have $M({\rm DG})/M({\HI}) = 0.23$ and 
$M({\rm DG})/M({\rm H^{\rm CO}_2}) = 1.24$, respectively.}
The values of those quantities thus derived are presented in Table~1.
Our estimates of DG mass are lower than those  found by \citet{Planck2011b} 
but are consistent with those of  \citet{Paradis2012}. 
{The estimated  DG mass is about  19 per cent
of the total gas mass}, very close to the value of 22 per cent of 
\citet{Planck2011b} and the value of 16 per cent of \citet{Paradis2012}. 
The small differences amongst those estimates are however probably insignificant owing to 
the large uncertainties involved in the analyses. On the other hand, the 
``close'' agreement between the results from different analyses, which utilize different data set of 
differing spatial resolutions and sky coverage  probably indicate a 
``true universal'' fraction of DG mass relative to that of all  gas, 
$\sim 20$ per cent on large Galactic scale.

\begin{figure}
\centering
   \includegraphics[width=0.45\textwidth]{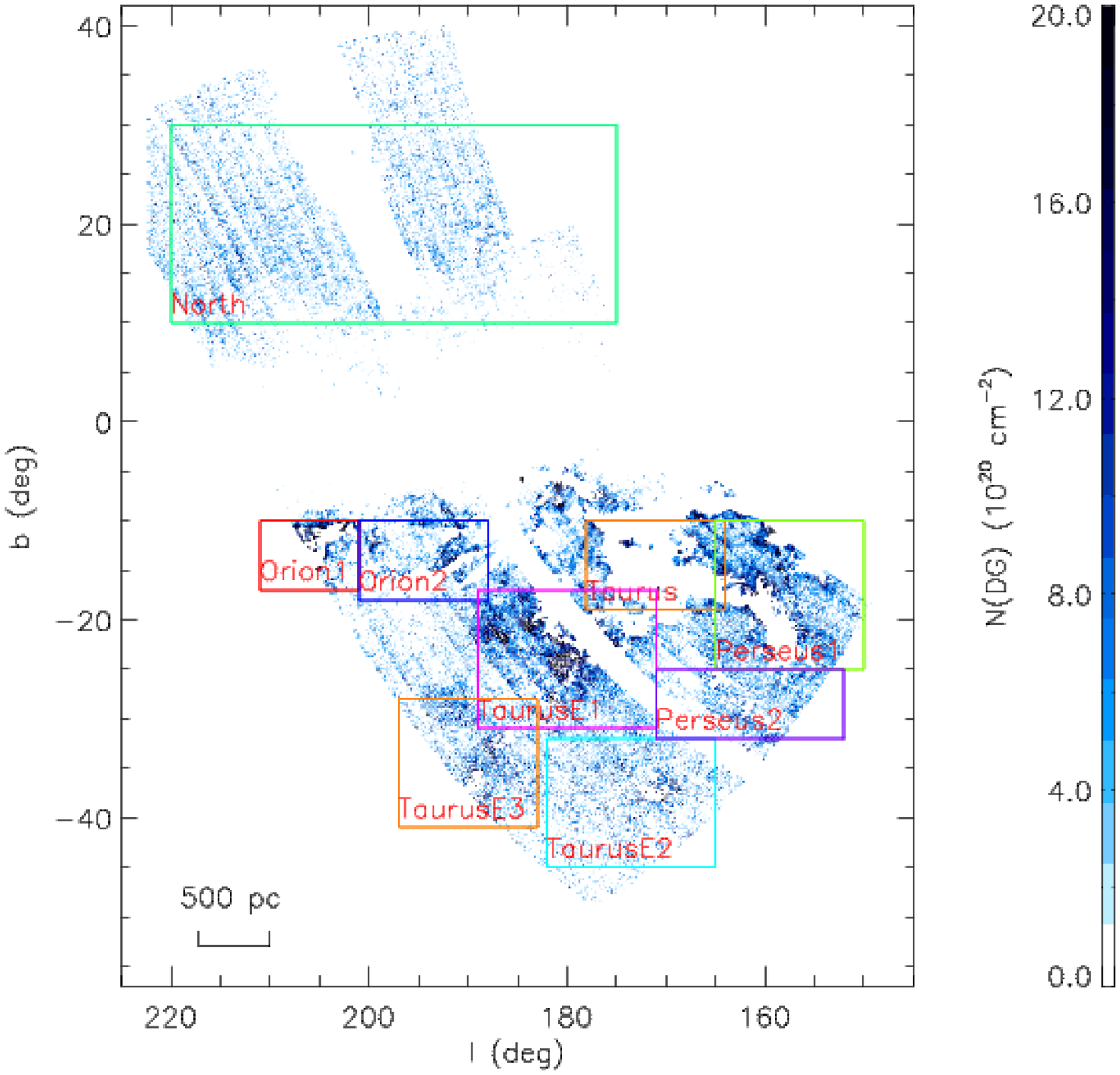}
   \includegraphics[width=0.45\textwidth]{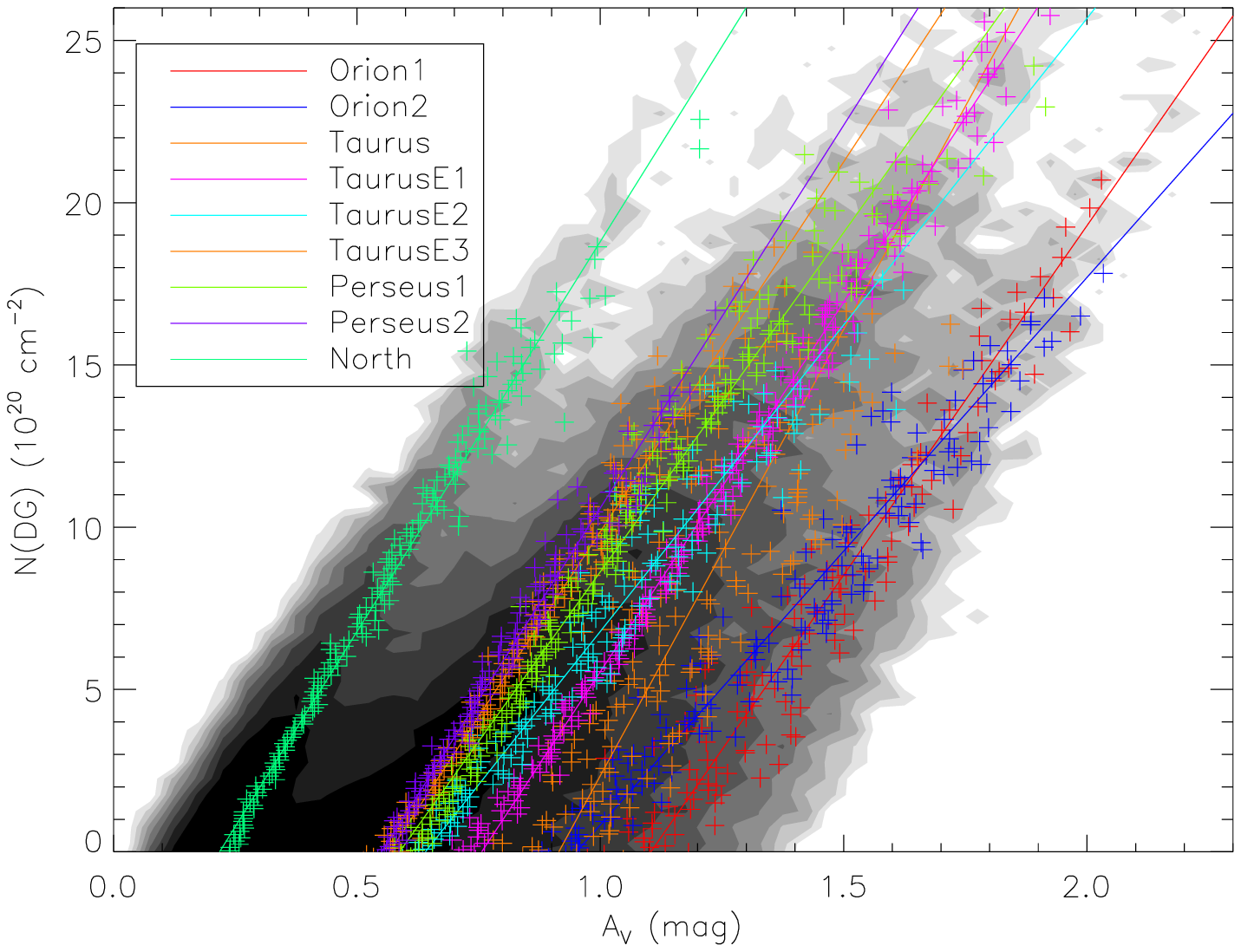}
\caption{ \em{Upper panel}\rm: Distribution of excess column densities. 
White areas are regions where 
no suitable extinction or H~{\sc i}  data are available,  
and regions where the CO emission is so  intense such that 
$W_{\rm CO} > 1\,\rm K\,km\,s^{-1}$. The squares delineate  the different
regions analyzed in the current work. {The map has an angular
resolution of 6\,arcmin.} \em{Lower panel}\rm: Correlation between the visual  extinction $A_V$ 
and the DG column density. Pluses of different colours are binned median
values for the  different regions, whereas  lines of different colours are  linear
fits to the binned median values.}
\label{darkg}
\end{figure}

The spatial distribution of DG derived from 
the current work is shown in the upper panel of Fig.~\ref{darkg} for the 
entire XSTPS-GAC footprint where the GALFA-HI data are available. 
Regions where $W_{\rm CO} > 1\,\rm K\,km\,s^{-1}$ have  
been excluded. The distribution  agrees well with that of \citet{Planck2011b} and 
clearly shows that the DG is distributed mainly around  
molecular clouds, such as the Taurus, Perseus and Orion clouds and 
exhibits features  similar to the dust distribution.
Significant amount of DG is found south of the Galactic plane 
within the footprint of XSTPS-GAC but little in the north. 
If we  compare the mass of DG to those of  H~{\sc i}  and CO amongst different clouds, 
we see significant variations. {For example, the ratio of 
$M({\rm DG})/M(\HI)$ varies from a value of 0.09 
for Orion\,1 to 0.22 for  the Taurus cloud and to 0.95 for the North region. Similarly, the ratio of 
$M({\rm DG})/M(\rm H^{\rm CO}_2)$ increases  from a value of 0.31 
for the Taurus cloud to 1.43 for  Taurus E1/Orion 2 and to 3.42 and 
4.01 respectively for Taurus E3 and E2.} The value of 
$M({\rm DG})/M(\rm H^{\rm CO}_2)$ for the North region is not listed in 
Table~1 given  the extremely large
uncertainty as a consequence of the lack of detection of CO emission in this region.
All the ratios quoted above are however highly dependent on the values of $DGR$ and
$X_{\rm CO}$ derived. The mass fraction of DG,  $f_{\rm DG}$, is defined as
$f_{\rm DG}=M({\rm DG})/[M({\rm DG})+M({\rm H^{\rm CO}_2})]$. Its values for the 
individual regions are also given in Table~1. {For the entire XSTPS-GAC 
region of $|b| > 10^\circ$, we find $f_{\rm DG}$ = 0.55,  in good agreement with 
the value of 0.54 found by \citet{Planck2011b} for the solar neighborhood, 
and comparable to the value  of 0.43 found by 
\citet{Paradis2012} for outer Galaxy of $|b|>10\degr$.
Our results are also in qualitative agreement
with the analysis by \citet{Grenier2005}, \citet{Abdo2010} and 
\citet{ Ackermann2011}, who find
that the DG component amounts to 40$-$400 per cent
of the CO-traced mass in local small molecular
clouds, such as the Cepheus, Polaris, Chamaleon clouds etc. 
\citet{Madden1997} find additional molecular mass in [C\,{\sc ii}]-emitting
regions equivalent to at least 100 times the mass of CO traced 
molecular gas in irregular galaxies. The values of $f_{\rm DG}$ in the current work are consistent with the 
theoretical work of \citet{Papadopoulos2002}, but
much larger than the value  modelled by \citet{Wolfire2010},
$f_{\rm DG}=0.3$, i.e. the DG mass is only about 43 percent of the CO traced molecular mass. 
The reason could be that the clouds we studied are all off the Galactic plane while 
\citet{Wolfire2010} consider the self-gravitating high-$A_V$ structures 
($A_V \ge 2$\,mag) that obey the Larson relations. By contrast, 
\citet{Papadopoulos2002} and other studies are concentrate on the diffuse-type
low-$A_V$ H$\rm_2$ clouds that are more likely to be CO-dark, 
i.e. the type of molecular ISM that is being studied in this work. 
\citet{Planck2011b} have also pointed out that \citet{Wolfire2010} adopted the 
\citet{Solomon1987} value for the mean column density of GMCs, which is 2-5 times larger than
the recent result of \citet{Heyer2009}.}
Amongst the different clouds, $f_{\rm DG}$ decreases from $\sim$ 0.7 in the Taurus extensions E1--E3
(with a relatively low average extinction) to $\sim$ 0.2 in the Taurus and 
Orion clouds (with a relatively high extinction on average). 
A linear correlation between the DG column density $N(\rm DG)$
and the visual  extinction $A_V$ is found for the different regions analyzed here
(lower panel of Fig.~\ref{darkg}). 
The ratios of those two quantities show little 
variations amongst  the different regions. We find that  $N(\rm DG)=(1.8 \sim 2.4) \times 10^{21}
(A_V-A^c_V)\,\rm cm^{-2}$, where  $A^c_V$ is a constant, representing the minimum  extinction
required for the presence of DG. The constant  $A^c_V$ thus represents the extinction of 
the H~{\sc i}-to-$\rm H_2$ transition region. Theoretically, this value changes with the
total H nuclei density $n$,
metallicity $z$, FUV field $G_0$ and $\rm H_2$ formation rate $R_f$ of the clouds 
\citep{Elmegreen1989, Elmegreen1993, Papadopoulos2002}. 
In the current work we find that its value  increases with 
the average extinction of the region, from  a value of 0.2\,mag for the 
North region to 1.1\,mag for Orion\,1.

\section{Summary}

In this paper, we have constructed  a 2D extinction map integrated out to a  distance of 4\,kpc 
for the footprint of  XSTPS-GAC using the data presented in Paper~I.
The map has a high angular resolution of 6\,arcmin and a low noise level  of 
0.18\,mag in $A_V$ [$\sim$60\,mmag in $E(B-V)$]. A comparison with the
widely used SFD dust map and the newly derived \citet{Schlafly2014} 
reddening map shows that our map has a high fidelity. 

The integrated extinction map is combined with high-resolution $\HI$ and
CO measurements to study the correlation between the interstellar gas content 
 and dust extinction within the XSTPS-GAC footprint.
The main results are:

\begin{enumerate}
\item {An  average value of dust-to-gas ratio $DGR=(4.15 \pm 0.01) \times
10^{-22}\, \rm mag\,cm^{2}$ and an average CO-to-H$_2$ conversion factor 
$X_{\rm CO} =(1.72 \pm 0.01) \times 10^{20}\,\rm 
cm^{-2}\,(K\,km\,s^{-1})^{-1}$ are found for the XSTPS-GAC footprint of $|b|>10\degr$.
The results are consistent with the findings of previous work.} 
\item We find a factor of $\sim$ 2 cloud-to-cloud variations in the values of
$DGR$ and $X_{\rm CO}$ amongst the molecular clouds south of the  
Galactic plane within the XSTPS-GAC footprint. 
Variations of $DGR$ could partly be caused by  the
differences in  dust properties, such as the grain size distribution of the 
different clouds.
{The $X_{\rm CO}$ values found for the molecular clouds are around $ \sim 1.5  \times 10^{20}\,\rm 
cm^{-2}\,(K\,km\,s^{-1})^{-1}$, in close agreement with the work of 
\citet{Polk1988}.}
\item We find a range of DG mass fraction similar to those found by \citet{
Planck2011b} and by \citet{Paradis2012}. {The mass of DG is about 23 and
124 per cent those of the atomic and the CO-traced molecular gas, 
respectively. The fraction of  molecular mass
of the  component, $f_{\rm DG}$, is about 0.55.}
The ``close'' agreement between results from different analyses, which utilize different data set of 
differing spatial resolutions and sky coverage  probably indicate a 
``true universal'' fraction of DG mass relative to that of all  gas, which
 is $\sim 20$ per cent on large Galactic scale.
Amongst the  different clouds, $f_{\rm DG}$ decreases with 
increasing  average extinction. Our DG mass estimates are much larger than
that modelled in \citet{Wolfire2010} but quite consistent with 
the theoretical work of \citet{Papadopoulos2002}.
In addition, we find that the DG column density 
has a linear relationship with the visual  extinction. {The average DG-to-dust ratio 
$N({\rm DG})/A_V$ is $\sim  2.0 \times 10^{21}\, \rm cm^{-2}\,mag^{-1}$. 
The extinction of the  H~{\sc i} -to-$\rm H_2$ transition layer varies between  0.22 -- 1.10\,mag,
and increases with increasing average  extinction of the cloud.} 

\end{enumerate}

\section*{Acknowledgements}

{We thank the anonymous referee for instructive comments that improved the 
current work significantly.
This work is partially supported by National Key Basic Research Program of China 
2014CB845700 and  China Postdoctoral Science Foundation 2014M560843.}

This work has made use of data products from the Guoshoujing Telescope (the 
Large Sky Area Multi-Object Fibre Spectroscopic Telescope, LAMOST). LAMOST
is a National Major Scientific Project built by the Chinese Academy of 
Sciences. Funding for the project has been provided by the National
Development and Reform Commission. LAMOST is operated and managed by the 
National Astronomical Observatories, Chinese Academy of Sciences.

This publication utilizes data from Galactic ALFA H~{\sc i}  (GALFA-HI) 
survey data set obtained with the Arecibo L-band Feed Array (ALFA) 
on the Arecibo 305m telescope. Arecibo Observatory is part of the 
National Astronomy and Ionosphere Center, which is operated by 
Cornell University under Cooperative Agreement with the U.S. 
National Science Foundation. The GALFA-HI surveys are funded 
by the NSF through grants to Columbia University, the University 
of Wisconsin, and the University of California.

\bibliographystyle{mn2e} 
\bibliography{dgr}

\end{document}